\documentclass[aps,prb,reprint,longbibliography,floatfix]{revtex4-2}
\usepackage[T1]{fontenc}
\usepackage[utf8]{inputenc}
\usepackage{lmodern}
\usepackage{amsmath,amssymb}
\usepackage{array}
\usepackage{graphicx}
\usepackage{xcolor}
\definecolor{linkblue}{RGB}{28,69,105}
\usepackage[colorlinks=true,linkcolor=linkblue,citecolor=linkblue,urlcolor=linkblue]{hyperref}
\hypersetup{pdftitle={Gallium at germanium surfaces: links to superconducting films and correlated adatom lattices},pdfauthor={Peter Jacobson},pdfsubject={Surface structure and superconductivity in gallium-doped germanium}}
\newcolumntype{L}[1]{>{\raggedright\arraybackslash}p{#1}}
\makeatletter
\g@addto@macro\@mkpream@relax{%
  \let\@endpbox\relax
  \let\do@row@strut\relax
  \let\ar@align@mcell\relax}
\AtBeginDocument{%
  \def\@hangfrom@section#1#2#3{\@hangfrom{#1#2}#3}%
  \def\@hangfroms@section#1#2{#1#2}}
\makeatother
\begin{document}
\title{Gallium at germanium surfaces: links to superconducting films and correlated adatom lattices}
\author{Peter Jacobson}
\affiliation{\mbox{School of Mathematics and Physics, The University of Queensland, Brisbane, Queensland 4072, Australia}\\
\normalfont\href{mailto:p.jacobson@uq.edu.au}{p.jacobson@uq.edu.au}}
\begin{abstract}
Epitaxial films grown on Ge(001) show that germanium hyperdoped with
substitutional Ga is superconducting. In earlier implanted films, Ga-rich
precipitates obscured the origin of the superconductivity. This review examines the sites Ga occupies on Ge(001) and Ge(111)
and how it enters the growing film. The Ga fraction expected from the growth fluxes is 17.9\%, yet the distribution of Ga among lattice sites is unknown. Comparisons include Ga on Si(001) and Si(111),
In, Sn, Pb, Sb and P on Ge(001), and elemental and two-dimensional Ga.
Where its surface coordination has been established, Ga has three bonds,
whereas substitutional Ga in the films is fourfold coordinated and acts
as an acceptor. Scanning tunnelling microscopy and X-ray standing waves
have assigned Ga sites on Ge(111), but the dilute Ga site on Ge(001), where the
films are grown, has yet to be measured. On Ge(111), Ga in the substrate may supply holes to the correlated Sn adatom lattice, as B supplies holes to the superconducting Sn lattice on Si(111). Whether Ga can supply holes depends on where it sits and whether it has three or four neighbours. On both surfaces, element-specific measurements and calculations comparing adsorption with substitution could settle the remaining site assignments.
\end{abstract}
\maketitle

\begingroup
\small
\makeatletter
\let\l@abstract\@gobbletwo
\let\l@subsection\@gobbletwo
\let\l@subsubsection\@gobbletwo
\renewcommand{\l@f@section}{\addvspace{1pt}}
\makeatother
\hypersetup{linktoc=all}
\tableofcontents
\endgroup

\section{Introduction}
\label{sec:introduction}

\begin{figure}[!t]
  \centering
  \includegraphics[width=\linewidth]{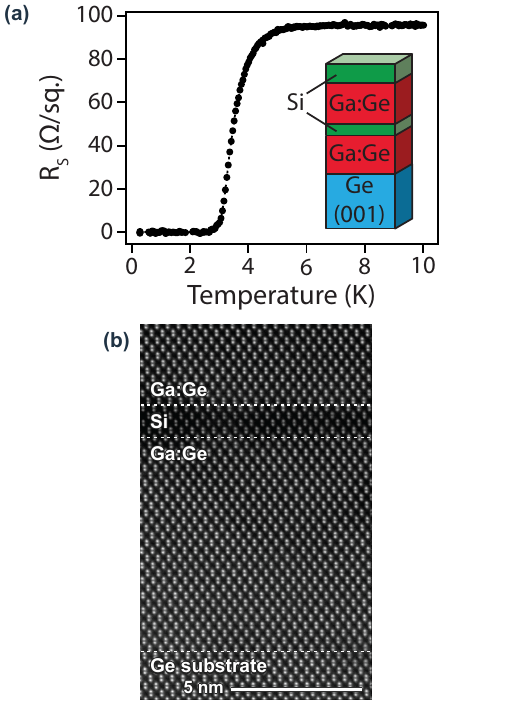}
  \caption{Superconductivity and film structure in epitaxial Ga:Ge on Ge(001).
(a) Sheet resistance through the transition of a Ga:Ge/Si/Ga:Ge trilayer. The inset shows the layer sequence.
(b) Cross-sectional high-angle annular dark-field scanning transmission electron microscopy (STEM) image showing the Ge substrate, the two Ga:Ge layers and the Si spacer. The dashed lines mark the interfaces. Scale bar, 5~nm. From Figs.~1(c) and 3(e) of Steele et al.~\cite{P32}.}
  \label{fig:fig01}
\end{figure}

In 1964 Cohen proposed that a semiconductor doped heavily enough could
become superconducting \cite{Cohen1964}. Superconductivity was found that
year in the degenerate semiconductors GeTe and SrTiO$_3$
\cite{Hein1964,Schooley1964}, four decades later in B-doped diamond and Si
\cite{Ekimov2004,Bustarret2006,P11}, and then in Ga-implanted Ge
\cite{Herrmann}. In Si and Ge the dopant concentrations
required for superconductivity lie far above the equilibrium solid solubilities
\cite{Trumbore1960}, and this work forms part of the wider resurgence of
interest in hyperdoping \cite{Tong2022,Sher2023}. Ga hyperdoping of Ge has
been pursued by implantation \cite{Heera2014,Prucnal,Sardashti}, chemical
vapour deposition \cite{P64,Ringwala2025} and molecular-beam epitaxy (MBE)
\cite{Strohbeen2023}. A persistent challenge has been to prepare
single-phase material, as implanted films often contain Ga-rich
clusters or precipitates \cite{Heera2014,Sardashti}, which leaves the origin
of the superconductivity uncertain. Steele et al.\ recently overcame this with
MBE-grown Ga:Ge films on Ge(001) that remain coherent with the substrate and
contain predominantly substitutional Ga, showing that Ge itself can be made
superconducting \cite{P32}. A Ga:Ge/Si/Ga:Ge trilayer, in which two Ga:Ge layers with a Ga concentration of 17.9\% expected from the growth fluxes are separated by a thin Si layer, has its
transition at 3.5~K (Fig.~\ref{fig:fig01}). Josephson junctions grown entirely by MBE
in this geometry, with the Si layer as a crystalline weak link, have
since been reported \cite{Knudsen2026}.

At these concentrations the isolated-dopant picture breaks down. How often
does a Ga atom have another Ga as a nearest neighbour? If Ga occupied the
sites of the diamond lattice at random, the fraction with at least one Ga
neighbour would be $1-(1-x)^4$ at a Ga fraction $x$, or 4\% at $x=0.01$
and 55\% at $x=0.179$. Neither the frequency of Ga neighbours relative to this estimate nor the variation in local composition across a film has been measured. The Ga distribution may also affect the hole density. In B-doped diamond,
calculations combined with secondary-ion mass spectrometry, transport and
Raman data indicate that B pairs form at high concentration and are
electrically inactive \cite{Bourgeois2006}, so the dopant distribution can change the carrier density. In the Ga:Ge trilayer, the Hall density of $4.15\times10^{21}$~holes~cm$^{-3}$ \cite{P32} corresponds to about one hole per two Ga atoms expected from the fluxes. The origin of this difference is unknown. The distribution of Ga may contribute.

The surfaces of Si and Ge can also support two-dimensional electronic
phases, from superconductivity in ultrathin Al films and single atomic
layers of Pb on Si(111) \cite{VanWeerdenburg2023Al,Zhang2010Monolayer} to
correlated phases in adatom lattices, which can depend on the carrier concentration of the substrate. One-third of a monolayer of Sn or Pb on
Si(111) or Ge(111) forms a triangular adatom lattice with a narrow,
nominally half-filled surface band \cite{Tejeda2007,Marchetti2025}. On
Si(111) the Sn lattice is a Mott insulator, and on heavily B-doped
substrates it is hole doped \cite{Ming2017} and becomes superconducting
\cite{Wu2020}, with evidence for chiral pairing \cite{Ming2023,Wu2026Chiral}.
On Ge(111) the Sn lattice is metallic as it cools through a structural
transition near 210~K. One Sn atom in three rises above the other two,
giving the surface its $(3\times3)$ periodicity
\cite{Zhang1999,Bunk1999SnGe}. Whether the lattice becomes a Mott insulator
on further cooling \cite{Cortes2006,Cortes2013} or remains metallic
\cite{Colonna2008,Morikawa2008} is still disputed
(Sec.~\ref{sec:snpb111}). Because Ga is a well-characterised acceptor in bulk Ge and several of its sites on Ge(111) are known, Ga-doped Ge is a natural substrate on which to test whether holes can be supplied to the
correlated Sn band, as they are by B in Si.

At the surface sites established on Si and Ge, Ga forms three bonds (Table~\ref{tab:gasites}). Substitutional Ga in the films has four
neighbours and is an acceptor (Sec.~\ref{sec:counting}). Equilibrium bulk
diffusion of Ga is negligible at the growth temperatures of the films
(Sec.~\ref{sec:growth}), so the fourth neighbour is gained, and the Ga
distribution set, at or just beneath the growth front. On Ge(111), scanning tunnelling microscopy (STM) supports a rest-atom assignment for dilute Ga \cite{MolinasMataZegenhagen1993} and X-ray standing waves locate Ga in the outer layer of dense Ga phases
\cite{P09}. On Ge(001), where the films are grown, the site of Ga at low
coverage has not been measured, and the available results are one
calculation \cite{Takeuchi} and studies of surfaces prepared at
400--500~$^\circ$C \cite{Gutakovsky1999,P15,Ji1997}.

This review collects what is known about Ga at Ge surfaces, beginning in
Sec.~\ref{sec:foundations} with the Ge(001) and Ge(111) surfaces, their
adsorption sites and the electron count that separates threefold from
fourfold Ga. Section~\ref{sec:ga001} covers Ga on Si(001), where its site
has been measured \cite{QianBedzyk1995,P99}, and Ga on Ge(001), and
Sec.~\ref{sec:comparators} the adsorbates In, Sn, Sb and P on Ge(001), which show when an adsorbate replaces Ge
\cite{FalkenbergIn1997,TomatsuSn,HofmannSn,MartinezGuerra2006SbAdsorption,Mattoni2013}.
Table~\ref{tab:adsorbates} also includes Pb. Section~\ref{sec:growth} considers Ga at the growth front of the films,
Sec.~\ref{sec:ga111} Ga on Ge(111) and the Sn adatom lattice, with Ga on
Si(111) for comparison \cite{Falta1996,P100}, and Sec.~\ref{sec:alphaga}
elemental and two-dimensional Ga. Section~\ref{sec:outlook} summarises the open questions (Table~\ref{tab:questions}). Pairing mechanisms in doped group-IV semiconductors and quantum
devices in Ge and Si lie outside the scope of this review and are covered
elsewhere \cite{P11,P55,Zwanenburg}.
\section{Ge surfaces and their adsorption sites}
\label{sec:foundations}

\begin{figure*}[tp]
  \centering
  \includegraphics[width=\linewidth]{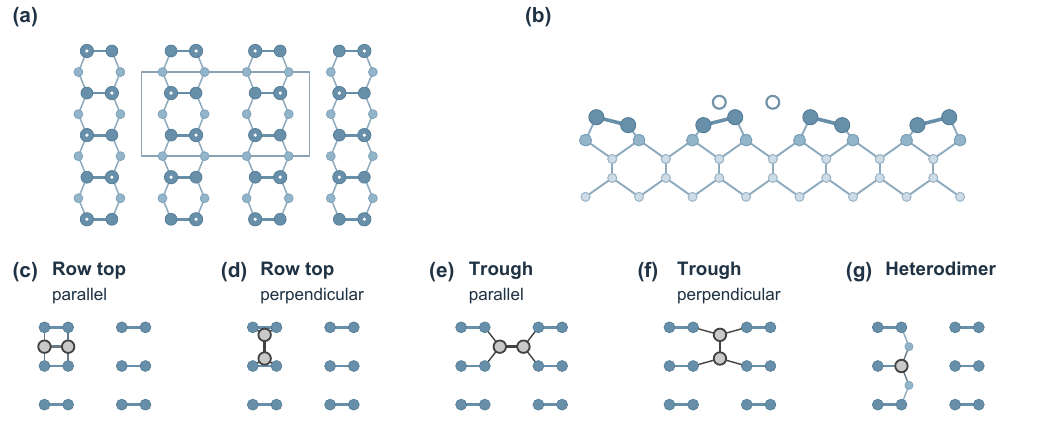}
  \caption{Ge(001) dimers and adsorption sites. (a) Plan view of the
$c(4\times2)$ surface. The thin solid rectangle outlines the conventional
cell. In the plan views the $[1\bar{1}0]$ dimer bond is horizontal and the
$[110]$ rows are vertical. White centres mark the upper atoms, and the
dimers buckle in opposite senses along and across the rows. The dimers in
(a,b) have a 2.45~\AA\ bond length and a $14^\circ$ tilt~\cite{P74}.
(b) Side view along $[110]$, showing one front dimer per row. Open circles
mark unoccupied ideal sites of the next Ge layer above the dimer centres
and trough midpoints. (c--f) Four possible geometries. Table~\ref{tab:adsorbates} gives the systems assigned to each. (c,d) Row-top ad-dimers parallel and perpendicular
to the Ge dimer bond, respectively~\cite{Zandvliet2000Dimers}. (e,f) Trough
ad-dimers parallel and perpendicular to the Ge dimer bond, respectively.
(g) An outer-layer heterodimer, with second-layer Ge included to show the
substituted atom's two back bonds. The substrate dimers are drawn symmetric
in (c--g). Blue Ge symbols lighten with depth from the dark outer layer
through the second layer to the pale, dark-outlined deeper atoms. Grey circles indicate generic adsorbates. Thick lines show dimer and ad-dimer bonds. Thin dark lines show
bonds from an added or substituted atom to the substrate, and thin grey
lines show substrate bonds, including the back bonds in (a). The substrate
Ge below the outer layer and the next-layer sites in (b) are at ideal bulk
positions for $a=5.658$~\AA.}
  \label{fig:fig02}
\end{figure*}

\begin{table}[tb]
\caption{Reference values for Si and Ge. Band gaps are at 300~K
\cite[Table~28.1, p.~566]{AshcroftMermin} and lattice constants at room temperature \cite{Hom}.
One monolayer (ML) is the atom density of a bulk-truncated surface layer.
On (111) a bilayer contains 2~ML. The adatoms on Ge(111)-$c(2\times8)$
occupy 0.25~ML, and a complete $\sqrt3$ adatom lattice occupies one-third
of a monolayer.}
\label{tab:scales}
\begin{ruledtabular}
\begin{tabular}{lrr}
Quantity & Si & Ge \\
\hline
Indirect band gap at 300~K (eV) & 1.12 & 0.67 \\
\hline
Lattice constant $a$ (\AA) & 5.431 & 5.658 \\
Bulk nearest-neighbour distance, $\sqrt3a/4$ (\AA) & 2.352 & 2.450 \\
Atoms per (001) layer, 1~ML ($10^{14}$~cm$^{-2}$) & 6.78 & 6.25 \\
Atoms per (111) layer, 1~ML ($10^{14}$~cm$^{-2}$) & 7.83 & 7.21 \\
(001) single-layer step, $a/4$ (\AA) & 1.358 & 1.415 \\
\end{tabular}
\end{ruledtabular}
\end{table}

\subsection{Ge(001)}
\label{sec:ge001}

\subsubsection{The clean surface}
\label{sec:clean001}

As on Si(001), the outer-layer atoms of Ge(001) pair into dimers, which line
up in rows, as shown in Fig.~\ref{fig:fig02}(a). Each dimer buckles, with a
mainly filled dangling bond on the upper atom and a mainly empty one on the
lower, and the buckling alternates along a row. Neighbouring rows buckle in phase in $p(2\times2)$ and in antiphase in $c(4\times2)$. The two reconstructions are nearly degenerate. At room temperature
$c(4\times2)$ domains coexist with $(2\times1)$ domains of rapidly flipping
dimers, which appear symmetric in STM \cite[Sec.~2]{P74}. Zandvliet reviews the clean
surface and its steps in detail \cite[Secs.~3.2 and 5.3]{P74}, and
Sec.~\ref{sec:gasteps} describes how Ga changes the steps.

\subsubsection{Adsorption sites}
\label{sec:sites001}

Adsorbates on Ge(001) either add to the surface in pairs or replace Ge in a
surface dimer. An adsorbed pair, or ad-dimer, sits on top of a dimer row,
above the Ge dimers, or in the trough between two rows, above the third Ge
layer \cite{Northrup001,Brocks1993}. In either place its bond is parallel or
perpendicular to the Ge dimer bonds, which gives the four sites of
Fig.~\ref{fig:fig02}(c--f), and parallel and perpendicular are used in this
sense throughout. The panels show possible geometries. Table~\ref{tab:adsorbates} identifies the assignments for each adsorbate and the evidence for them. Ge ad-dimers on Ge(001) are found both on the rows and in
the troughs \cite{Zandvliet2000Dimers}. A substituted atom instead takes the
place of one Ge atom of a surface dimer, forming the heterodimer of
Fig.~\ref{fig:fig02}(g), and the Ge it displaces goes into an ad-dimer, an
island or a step (Sec.~\ref{sec:comparators}).

The sites differ in their registry with the next Ge layer, that is, in
whether they continue the bulk lattice. The next layer would occupy the
positions marked in Fig.~\ref{fig:fig02}(b), above the centre of each dimer
and midway across each trough, and would dimerise along the rows of the
layer beneath. A perpendicular ad-dimer has that orientation and can sit in
registry, as some Sn pairs do \cite{TomatsuSn}, whereas a parallel ad-dimer
cannot. A substituted atom already occupies a lattice site of the outer
layer.

At low coverage the most common pair orientation differs between the
group-III and group-V adsorbates (Table~\ref{tab:adsorbates}). Indium forms parallel
trough pairs on Ge(001), as Ga does on Si(001), whereas Sb and P form
perpendicular pairs on top of the rows
\cite{FalkenbergIn1997,QianBedzyk1995,P99,MartinezGuerra2006SbAdsorption,Mattoni2013}.
Of the group-IV adsorbates, Sn forms pairs in both orientations and already
replaces some Ge in dimers, whereas the half-monolayer Pb phase contains parallel trough pairs over
intact Ge dimers \cite{TomatsuSn,HofmannSn,Takeuchi2002Pb}. With rising
coverage the pairs order into $2\times2$ arrangements, and beyond 0.5
monolayers (ML, defined in Table~\ref{tab:scales}) structures appear that
break Ge dimers or reshape the surface
\cite{QianBedzyk1995,Takeuchi2002Pb,FalkenbergIn1997,FalkenbergIn2002}.

\begin{figure*}[tp]
  \centering
  \includegraphics[width=\linewidth]{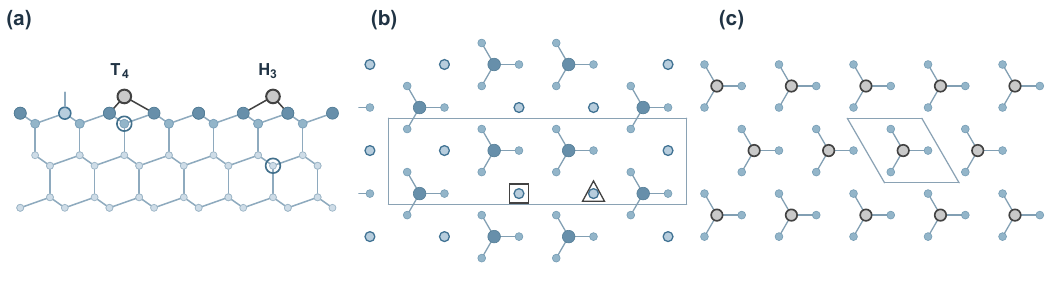}
  \caption{Sites and adatom lattices on Ge(111). (a) Side view along $[01\bar{1}]$. The rings mark the substrate atoms beneath the $T_4$ and $H_3$ hollows, in the second and fourth atomic layers, respectively. The short line on the left marks a rest-atom dangling bond. (b) The $c(2\times8)$ reconstruction, after Molinàs-Mata and Zegenhagen~\cite{MolinasMataZegenhagen1993}. The rectangle outlines one cell. The square and triangle mark rest atoms with four and three adatom neighbours, respectively. (c) The one-third-monolayer $(\sqrt3\times\sqrt3)R30^\circ$ $T_4$ adatom lattice at the same scale, with one unit cell outlined. In (b,c), $[\bar{2}11]$ is horizontal and $[01\bar{1}]$ is vertical. In (b,c), large blue circles are Ge adatoms, pale outlined circles are rest atoms and small circles are outer-layer Ge bonded to an adatom. In (a), Ge symbols lighten with depth as in Fig.~\ref{fig:fig02}. Grey circles are generic adatoms. The substrate is drawn at ideal bulk positions for $a=5.658$~\AA. Adatom heights in (a) are schematic.}
  \label{fig:fig03}
\end{figure*}

\begin{table*}[tb]
\caption{Structures reported for Ga and other adsorbates on Ge(001),
from low coverage at low temperature to higher coverage or after
annealing, with the destination of displaced Ge and the evidence behind
each entry. Panel letters refer to the geometries in Fig.~\ref{fig:fig02}. Parallel and perpendicular describe the adsorbate bond relative to the substrate dimer bond. Roman numerals give the
periodic-table group. XSW, X-ray standing waves; DFT, density
functional theory; LEED and RHEED, low-energy and reflection
high-energy electron diffraction.}
\label{tab:adsorbates}
\footnotesize
\setlength{\tabcolsep}{4pt}\renewcommand{\arraystretch}{1.18}
\begin{ruledtabular}
\begin{tabular}{L{0.06\textwidth}L{0.21\textwidth}L{0.26\textwidth}L{0.18\textwidth}L{0.20\textwidth}}
 & Low coverage, low temperature & Higher coverage or after annealing & Displaced Ge & Evidence \\
\hline
Ga (III) & Dilute site undetermined on Ge(001). On Si(001), measured parallel trough pairs (e), with $2\times2$ order at 0.3~ML & On Ge(001), DFT favours (e) over (f) in an ordered $2\times2$ half-monolayer; about 0.5~ML deposited at 400--500~$^\circ$C converts double-layer steps to single-layer steps; after annealing near 500~$^\circ$C, $7\times5.5$ mounds with 0.3--1~ML retained, and only $(2\times1)$ LEED below 0.3~ML & Ge dimers proposed at mound tops, with their origin unresolved & Si: XSW, STM and DFT. Ge: DFT; steps from RHEED; mounds from STM and bonding rules \cite{QianBedzyk1995,P99,Takeuchi,Gutakovsky1999,P15} \\
In (III) & Parallel trough pairs (e) in two-dimensional islands from 0.03~ML & Above 200~$^\circ$C, mixed In--Ge $3\times4$ and $4\times4$ units ($7\times4$) up to 0.5~ML; near 1~ML, $5\times4$ and then $\{103\}$ facets at 350--500~$^\circ$C & Four Ge per mixed unit; in islands after about 200~$^\circ$C, at steps after about 400~$^\circ$C & STM at positive and negative sample bias and DFT; amount from island areas \cite{FalkenbergIn1997,FalkenbergIn2002} \\
Sn (IV) & At room temperature, STM resolves chains along $\langle310\rangle$ and $\langle110\rangle$, assigned to parallel trough dimers (e); epitaxial triplets have one perpendicular row-top and two perpendicular trough dimers (d,f). Sn--Ge heterodimers (g) also form & Heterodimers increase after 147~$^\circ$C; at 1.24~ML the first layer is closed ($<2$\% Ge) and higher layers have started & To steps or into mixed ad-dimers & STM at positive and negative sample bias and DFT; mixed ad-dimers a possible configuration \cite{TomatsuSn,Tomatsu2007Science,HofmannSn} \\
Pb (IV) & Not reported & 0.5~ML: $(2\times2)$ layer of parallel trough pairs (e) over intact Ge dimers. 0.75~ML: $c(4\times8)$ with Pb dimer chains and half the Ge dimers broken & None in the $(2\times2)$ layer; not reported at higher coverage & Bias-dependent STM and DFT \cite{Takeuchi2002Pb} \\
Sb (V) & Perpendicular row-top Sb$_2$ (d), with Sb$_4$ clusters and flat tetramers, at room temperature & Incorporation into the top layer from 247~$^\circ$C; pits and Ge islands after 417~$^\circ$C & Ge islands; calculated exchange favourable only with the Ge at bulk energy & Occupied-state STM and DFT \cite{MartinezGuerra2006SbAdsorption} \\
P (V) & Near 100~$^\circ$C, majority row-top feature assigned to perpendicular P$_2$ (d, 86\%); minority row-top feature tentatively assigned to parallel P$_2$ (c, 11\%) & 1~s at 250~$^\circ$C gives Ge--P heterodimers (g); near 1~ML, 5~min at 360~$^\circ$C gives parallel P--P dimers & Elongated islands after 250~$^\circ$C; single-layer steps after 360~$^\circ$C & STM at positive and negative sample bias, assigned by analogy with P/Si(001) \cite{Mattoni2013} \\
\end{tabular}
\end{ruledtabular}
\end{table*}

\subsection{Ge(111)}
\label{sec:ge111}

\subsubsection{The clean surface}
\label{sec:clean111}

Along [111] the diamond lattice is a stack of bilayers, and each outer atom
of a bulk-truncated (111) surface has three bonds into its bilayer and one
dangling bond \cite{Monch}. An adatom sits in a hollow between three
outer-layer atoms, bonds to all three and keeps one dangling bond of its
own. The two hollows are shown in Fig.~\ref{fig:fig03}(a). In the $T_4$
hollow the adatom lies directly above a second-layer atom, which is close
enough to count as a weak fourth neighbour, and in the $H_3$ hollow it lies
above a fourth-layer atom, with nothing beneath it in the outer bilayer
\cite{Monch}. Outer-layer atoms that are not bonded to an adatom keep their
dangling bonds and are called rest atoms.

Annealed Ge(111) forms the $c(2\times8)$ reconstruction of
Fig.~\ref{fig:fig03}(b), which is built from adatoms and rest atoms alone.
Ge adatoms occupy $T_4$ sites at one-quarter of a monolayer (Table~\ref{tab:scales}), so that
three-quarters of the outer-layer atoms bond to an adatom and one-quarter
remain as rest atoms. The number of dangling bonds thereby falls from one to
one-half per $1\times1$ cell. The adatoms are arranged in rows, and the
centred $2\times8$ cell, which contains four adatoms and four rest atoms, is
built from alternating $2\times2$ and $c(4\times2)$ blocks
\cite{MolinasMataZegenhagen1993,MolinasMata1993Doped}. Electrons transfer
from the adatom dangling bonds to those of the rest atoms, so the rest-atom
dangling bonds are largely filled and the adatom dangling bonds largely
empty \cite{Monch,P41}. Feenstra et al.\ resolved both states by tunnelling
spectroscopy at low temperature \cite{P41}. The empty adatom band begins 0.5~eV above the bulk valence-band maximum, once tip-induced band bending is allowed for. At low tunnelling current the filled rest-atom peak lies about 0.9~eV below the bulk valence-band maximum. STM accordingly images the rest atoms at negative sample bias and the adatoms at positive bias. On p-type crystals at low temperature the contrast reverses below about $-2$~V, when tip-induced band bending brings the adatom states below the Fermi level and they fill \cite{P41}. Above about 300~$^\circ$C Ge(111) loses the
$c(2\times8)$ order, and Ga can stabilise the high-temperature phase at room
temperature \cite{P12}.

Si(111), by contrast, forms the $7\times7$ dimer-adatom-stacking-fault
reconstruction, whose cell contains 12 adatoms, nine dimers along the sides
of its two triangular halves and a stacking fault in one half
\cite{Takayanagi1985,Monch}. The two clean reconstructions share adatoms and
rest atoms, but only Si(111)-$7\times7$ also changes the stacking and
bonding of the outer bilayer. Ga sites on the two surfaces are therefore
compared through the bulk lattice beneath them, as $T_4$, $H_3$ or
outer-layer sites, and not through the clean reconstructions.

\subsubsection{Adsorption sites and adatom lattices}
\label{sec:sites111}

On Ge(111) the adsorbates considered in this review take sites as single
atoms, either in the same hollows as the Ge adatoms or in place of an atom
of the outer layer, and the pairs, troughs and orientations of Ge(001) have
no counterpart. A substituted atom replaces a rest atom, or an atom in the
outer half of the bilayer within a denser adsorbate phase, and is then
bonded to three atoms below it \cite{MolinasMataZegenhagen1993,P09}. It
occupies a lattice site, whereas neither hollow continues the bulk lattice,
in which the next atom sits directly above an outer-layer atom.

A $(\sqrt3\times\sqrt3)R30^\circ$ adatom lattice forms on both Si(111) and
Ge(111), despite their different clean reconstructions, because it needs
only an unreconstructed outer bilayer beneath it. With one-third of a
monolayer of adatoms on $T_4$ sites, every outer-layer atom is bonded to an
adatom, as in Fig.~\ref{fig:fig03}(c). No rest atoms remain, and the only
dangling bonds are those of the adatoms, one for every three $1\times1$
cells. One-third of a monolayer of
Sn or Pb forms this lattice on Ge(111), and Sn also forms it on Si(111)
\cite{Carpinelli1996,Bunk1999SnGe,Tejeda2007,Ming2017}. Ga forms the $T_4$
lattice on Si(111) \cite{Falta1996,P100}, but no corresponding
one-third-monolayer Ga adatom lattice has been reported on Ge(111)
\cite{P12} (Sec.~\ref{sec:ga111ge}).

Before the lattice can form on Ge(111), the Ge adatoms of the
$c(2\times8)$ cell have to leave their sites, and a short anneal after
room-temperature deposition is enough, at about 250~$^\circ$C for Pb
\cite{Carpinelli1996} and about 290~$^\circ$C for Sn \cite{Halbig2019}.
 On Si(111) the dimers and stacking fault of the $7\times7$ cell must also
be removed, and the Sn lattice is annealed at about 700~$^\circ$C
\cite{Halbig2019}. Some Ge remains in the adatom lattice in place of Sn
\cite{Melechko2000}, and below one-third of a monolayer of Pb a mosaic phase
forms with Ge in place of the missing Pb \cite{Mascaraque1998}.
In the Sn lattice these Ge atoms pin the $(3\times3)$ order (Sec.~\ref{sec:snpb111}).

\subsection{Electron counting}
\label{sec:counting}

For Ga with three covalent bonds, formal electron counting assigns
one Ga valence electron to each bond. This leaves the dangling-bond orbital
nominally empty \cite{Monch}. Its actual occupation depends on hybridisation
with substrate states and on the Fermi level. On Si(001), for example,
tunnelling spectroscopy resolves an unoccupied state attributed to the Ga
pair (Sec.~\ref{sec:gasi001}).
Where the geometry allows, threefold Ga relaxes towards planar bonding, as substituted
Ga does in dense Ge(111) phases \cite{Monch,P09}. In a Ga pair on Ge(001),
the two Ga atoms and the four Ge atoms they bond to supply ten electrons,
enough to fill the five bonding states in this count
\cite{Monch,Northrup001}. A Sn or Pb atom with the same coordination keeps
one electron in its dangling orbital, the origin of the nominally
half-filled band of the $\sqrt3$ adatom lattices, and threefold Sb or P
keeps a lone pair \cite{Monch,Marchetti2025}. At a fourfold substitutional site in Ge, Ga supplies one fewer electron
than the Ge it replaces. An isolated Ga acceptor leaves a hole in the
valence band when it ionises. The threefold surface site has no such
one-electron deficit in its bonds.
\section{Ga on Si(001) and Ge(001)}
\label{sec:ga001}

\begin{figure*}[tp]
  \centering
  \includegraphics[width=\linewidth]{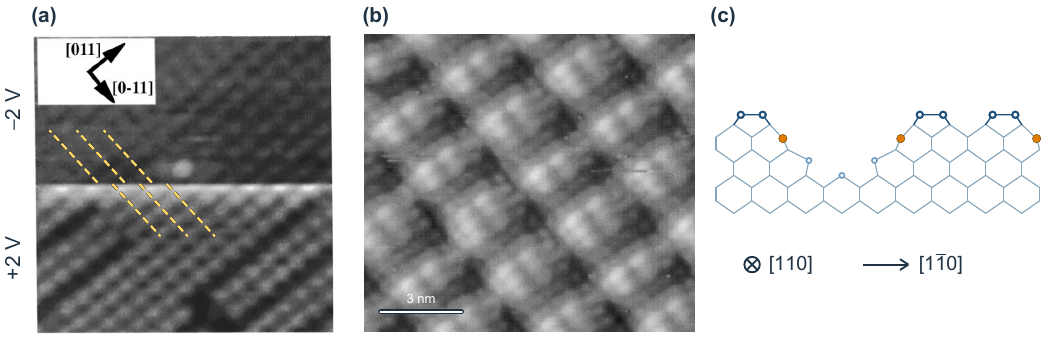}
  \caption{Ga on Si(001) and Ge(001). (a) STM of the $2\times2$ Ga/Si(001) surface. The sample bias was switched midway through the scan from $-2.0$~V (upper half) to $+2.0$~V (lower half) at 1.0~nA. The dashed guides mark three neighbouring rows in each half. The lower set is offset schematically to show the half-period shift along the Si dimer rows reported by Sakama et al. From Fig.~1(a) of Sakama et al.~\cite{P99}. (b) Empty-state STM of the annealed $7\times5.5$ Ga/Ge(001) phase (1.4~V, 1.0~nA), showing rows of mounds separated by deep troughs. (c) Section through a deep trough in the structure proposed by Gai et al. In the domain drawn, $[1\bar{1}0]$ points to the right along the top Ge dimer bonds and $[110]$ points into the page. Orange circles are substituted Ga and open centres denote Ge with a dangling bond. The Ga positions follow from the proposed bonding geometry. STM image from Fig.~3(d) of Gai et al.~\cite{P15} and section redrawn after their Fig.~4(b).}
  \label{fig:fig04}
\end{figure*}

\subsection{Ga on Si(001) for comparison}
\label{sec:gasi001}

On Si(001), the only (001) surface on which the site of Ga has been
measured, Ga at low coverage forms ad-dimers in the troughs between the
substrate dimer rows, with the Ga--Ga bond parallel to the Si dimer bonds
\cite{P99,QianBedzyk1995,Northrup001}. X-ray standing waves (XSW) use the
standing wave formed by a Bragg reflection to measure the height of a chosen
element above a set of bulk lattice planes, and combining reflections gives
its lateral site. For a room-temperature surface with an average of 0.3~ML
Ga and $2\times2$ order, XSW gives a Ga--Ga separation of
$2.58\pm0.04$~\AA\ \cite{QianBedzyk1995}. STM registers the pairs to the
substrate rows at positive and negative sample bias \cite{P99}, as in
Fig.~\ref{fig:fig04}(a), and density functional theory (DFT) places the
parallel pair below the perpendicular one in energy \cite{Northrup001}. Each
Ga makes two bonds to Si and one to its partner (Sec.~\ref{sec:counting}),
and tunnelling spectroscopy shows an empty state attributed to the remaining
Ga orbital \cite{P99}, which is the STM signature of threefold Ga on Si(001).

The pairs do not stay isolated but line up in rows perpendicular to the
substrate dimer rows, packing into a complete $2\times2$ layer near half a
monolayer \cite{EvansNogami1999,P97,Takeuchi}. Short rows predominate. In the growth model, Ga can attach to an existing row only at its ends, and new rows continue to nucleate at defects and on the open terrace \cite{Albao2005,Albao2006Reply}.
Calculations for Al/Si(001) show how a pair added at a row end creates a new
reactive end, so rows grow by attachment at their ends \cite{Brocks1993}.

\subsection{Calculated Ga pairs on Ge(001)}
\label{sec:gadilute}

In Takeuchi's DFT calculation for an ordered half-monolayer of Ga on Ge(001), the parallel trough pair [Fig.~\ref{fig:fig02}(e)] is lower in energy than the perpendicular pair [Fig.~\ref{fig:fig02}(f)] by 0.63~eV per pair \cite{Takeuchi}. The corresponding differences for Al and In are 0.65 and 0.37~eV. The perpendicular pair
stretches the Ge dimers it bonds to from a calculated 2.47~\AA\ to 2.94~\AA, against 2.59~\AA\ for the parallel pair, and Takeuchi
attributes the energy difference to this weakening of the Ge dimers
\cite{Takeuchi}. On Si(001) DFT gives the same ordering with a larger
difference, 1.32~eV per pair \cite{Northrup001}.
 The parallel orientation is the one In takes on Ge(001) \cite{FalkenbergIn1997}.

The calculation does not treat isolated pairs, Ga on top of a row or Ga
substituted into a Ge dimer \cite{Takeuchi}. The parallel pair it favours is
not in registry with the next Ge layer, whereas the higher-energy
perpendicular pair sits where a next-layer dimer would
(Sec.~\ref{sec:sites001}). Ga buried as a parallel pair would therefore not be on lattice sites, and the pair must rearrange before the Ga becomes substitutional. No STM study has located Ga on Ge(001)
at low coverage, so the site of dilute Ga has not been determined by
experiment.

\subsection{Ga-induced step changes on Ge(001)}
\label{sec:gasteps}

Step edges, where atoms have lower coordination than on a terrace, are
natural sites for adsorbates to attach, and Ga changes the step structure of
both Ge(001) and Si(001). Clean vicinal Ge(001) has single-layer steps at
small miscut angles and mainly double-layer steps above 5--6$^\circ$
\cite[Secs.~3.2 and 5.3]{P74}. On Ge(001) miscut by 5$^\circ$, reflection high-energy electron diffraction (RHEED) showed double-layer steps before Ga deposition \cite{Gutakovsky1999}. On this surface, about half a
monolayer of Ga deposited at 400--500~$^\circ$C converts the double-layer
steps into single-layer steps \cite{Gutakovsky1999}. The surface was a Ge
film grown on GaAs(001), and RHEED along the step edges showed the step height halving from $a/2$ to
$a/4$, whereas exposure to As$_4$ alone left the double-layer steps intact
\cite{Gutakovsky1999}. The step conversion was examined only at 0.5~ML, so the coverage at
which the conversion begins is not known.

On vicinal Si(001) miscut by 4$^\circ$, Nogami et al.\ followed the same
change by STM \cite{P98}. Ga deposited at room temperature leaves the
double-layer steps unchanged. After deposition or annealing at
450~$^\circ$C the double-layer steps kink or split into single-layer steps,
increasingly with Ga dose from below 0.1~ML, and by 0.4~ML many have split
and $2\times2$ arrays of Ga pairs cover much of the terraces. Below 0.1~ML
no Ga is seen on the terraces, and Nogami et al.\ suggest that it is taken
up at the step edges. On well-oriented Si(001), annealed Ga leaves the
single-layer steps unaffected \cite{P98}.

\subsection{The annealed \texorpdfstring{$7\times5.5$}{7x5.5} phase and its proposed model}
\label{sec:gamounds}

Annealed submonolayer Ga reconstructs Ge(001) into a single phase, reported
in three studies \cite{Ji1997,Gutakovsky1999,P15}, and it is the only structure in which Ga has been proposed to occupy substituted sites of
Ge(001). Ji et al.\ list it as a $(7\times7)$ low-energy electron
diffraction (LEED) pattern after slow-desorption anneals, and note that Ga,
unlike In, does not facet Ge(001) to $\{103\}$ \cite{Ji1997}. Gutakovsky et
al.\ report a
$(7\times7)$ RHEED pattern on vicinal and well-oriented Ge(001) after
about 1~ML of Ga \cite{Gutakovsky1999}, which is probably the same phase.
Gai et al.\ show by LEED that the structure is incommensurate and by STM
that it consists of rectangular mounds seven substrate units long and five
or six wide, in straight rows separated by deep troughs, as in
Fig.~\ref{fig:fig04}(b). They identify it with the $(7\times7)$ of Ji et
al.\ \cite{P15}. After deposition of 1~ML or more and about ten minutes near
500~$^\circ$C, this is the only reconstruction for retained coverages
between 0.3 and 1~ML, and it coexists with islands that Gai et al.\
attribute to surplus Ga \cite{P15}. Surfaces retaining less than about
0.3~ML show only the $(2\times1)$ LEED pattern of clean Ge(001) \cite{P15},
which leaves open whether that Ga sits in disordered pairs, at steps or
beneath the surface.

Gai et al.\ place four or six Ge dimers on top of each mound, with Ga atoms
below them in substituted and added positions, each Ga making three
sp$^2$-like bonds and keeping no dangling bond \cite{P15}. Figure~\ref{fig:fig04}(c) shows the proposed structure at a deep trough. The model was built to satisfy rules that the
same group drew from its LEED, Auger electron spectroscopy and STM studies
of group-III metals on Si and Ge. In these rules the metal atoms form
sp$^2$-like bonds, take substituted sites where they create only compressive
stress and added sites only where they create no tensile stress, and leave
the surface with some dangling bonds to relieve stress \cite{Ji1997,P15}.
The group lists its Ga/Ge(113) structures among those that obey the rules
\cite{Ji1997,GaiGe113}. Assuming equal numbers of $7\times5$ and $7\times6$
mounds, the model contains 0.39~ML Ga and 0.6 dangling bonds per $1\times1$
cell, against one on clean Ge(001) \cite{P15}. Because the islands hold part of the Ga, the retained coverage need not equal the Ga content of the mounds.

The model matches the features seen in STM, namely the outline of the mounds
and, through a mixture of mounds five and six units wide, the average period
of 5.5 \cite{P15}. It takes the STM contrast to be topography and assigns
the elements from the bonding rules, since STM does not identify the element
of an imaged atom, as Gai et al.\ note \cite{P15}. The model has not been
relaxed or tested by DFT.

\subsection{Aggregation of Ga on (001) surfaces}
\label{sec:gadistribution}

\begin{figure*}[tp]
  \centering
  \includegraphics[width=\linewidth]{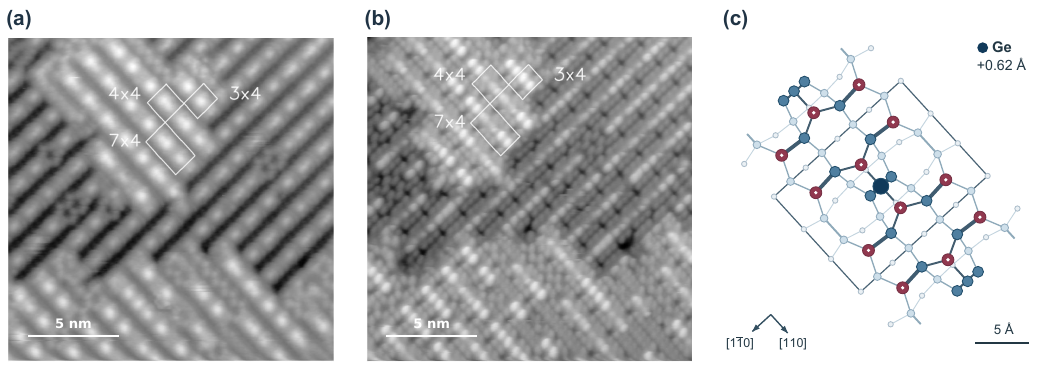}
  \caption{Mixed In--Ge units on Ge(001). (a,b) STM of the same area at $-1.8$ and $+1.8$~V sample bias, respectively, at 0.9~nA. The outlines mark $3\times4$, $4\times4$ and $7\times4$ cells. Scale bars, 5~nm. (c) Plan view of the relaxed $3\times4$ structure from Tables I and II of Falkenberg et al.~\cite{FalkenbergIn2002}. The outlined cell has the orientation of the $3\times4$ cell in (a). Half a neighbouring cell is shown at each end along the dimer rows. White-centred burgundy circles are In. Outer-layer Ge is blue, and the two underlying Ge layers are successively paler. Thick bonds mark surface dimers, including mixed In--Ge dimers. The large dark Ge atom between the rows lies 0.62~\AA\ above its two In neighbours and gives the occupied-state maximum. Scale bar, 5~\AA. STM images from Falkenberg et al., Fig.~1(a,b)~\cite{FalkenbergIn2002}.}
  \label{fig:fig05}
\end{figure*}

The superconducting films are grown at 95--150~$^\circ$C
\cite[Supplementary Fig.~S3]{P32}, and none of the results for Ga on Si(001)
or Ge(001) was obtained in that range. The experiments on Ge(001) were made on surfaces prepared at 400--500~$^\circ$C (Secs.~\ref{sec:gasteps} and \ref{sec:gamounds}). The proposed model for the $7\times5.5$ mounds formed at these higher temperatures places Ga in threefold sites. The Ga in the films is fourfold coordinated.

Wherever Ga could move in these experiments it aggregated, into rows and
$2\times2$ domains on Si(001) \cite{P99,EvansNogami1999,Albao2005}, at step
edges on vicinal Si(001) in Nogami et al.'s interpretation \cite{P98}, and
into mounds and islands on annealed Ge(001) \cite{P15}. Indium likewise forms
two-dimensional islands of pairs on Ge(001) from 0.03~ML
\cite{FalkenbergIn1997}. This aggregation makes the composition uneven across rows, domains and islands in static layers below saturation coverage. Within each pair, Ga already has a Ga neighbour. The pair must rearrange before its atoms can occupy substitutional sites (Sec.~\ref{sec:gadilute}), so the surface observations do not fix the neighbour statistics in the film. The reference for the films remains the random alloy, in which 55\% of the Ga atoms have
a Ga neighbour at 17.9\% Ga (Sec.~\ref{sec:introduction}). If Ga aggregates
in the same way on Ge(001) at growth temperatures, the films could by
analogy contain locally Ga-rich regions.
\section{Other adsorbates on Ge(001): substitution and displaced Ge}
\label{sec:comparators}

Between the growth surface and the film, Ga must change from an adsorbed
atom into a substituted one, and this change has not been observed for Ga on
Ge(001). It has been followed for In, Sn, Sb and P, from groups III, IV and
V (Table~\ref{tab:adsorbates}).

\subsection{Indium (group III): mixed In--Ge units and the Ge they release}
\label{sec:in001}

Indium, the next group-III element after Ga, is the adsorbate most like it,
and on Ge(001) it has been studied in more detail than Ga. STM has followed it from
0.03~ML to about a monolayer, at room
temperature and after annealing, and the annealed structures have been
modelled by DFT \cite{FalkenbergIn1997,FalkenbergIn2002}. The annealed In structures differ from the $7\times5.5$ phase that Ga forms (Sec.~\ref{sec:gamounds}).

At room temperature In forms parallel trough pairs
(Sec.~\ref{sec:sites001}). Annealing above 200~$^\circ$C converts them into
mixed In--Ge units with $3\times4$ and $4\times4$ cells, shown in
Fig.~\ref{fig:fig05}(a,b) \cite{FalkenbergIn1997,FalkenbergIn2002}. The
relaxed models, which were matched to STM images at positive and negative sample bias,
place indium in two Ge dimers and an In--Ge--In group between the rows. The
central Ge atom of that group gives the occupied-state maximum of
the $3\times4$ unit [Fig.~\ref{fig:fig05}(c)] \cite{FalkenbergIn2002}. With the displaced Ge at the
energy of bulk Ge, the mixed units lie 0.07~eV per $1\times1$ cell below the
layer of pairs. Falkenberg et al.\ conclude that the pair layer is
metastable, and that it forms first because it leaves the Ge dimers intact
and displaces no Ge \cite{FalkenbergIn2002}.

From the In pairs counted before annealing and the area of new Ge islands
afterwards, Falkenberg et al.\ estimated that each mixed unit releases four
Ge atoms, against five or six in the relaxed models
\cite{FalkenbergIn1997,FalkenbergIn2002}. The
displaced Ge forms terrace islands after annealing near 200~$^\circ$C. After
annealing near 400~$^\circ$C the islands have gone while the density of
mixed units is unchanged, which Falkenberg et al.\ explain by the Ge
reaching steps only at the higher temperature \cite{FalkenbergIn1997}.

\subsection{Sn (group IV): substitution during room-temperature deposition}
\label{sec:groupIV}

Sn is the only adsorbate in Table~\ref{tab:adsorbates} that replaces Ge
without annealing. During deposition at room temperature some Sn atoms enter
the surface directly, each taking the place of one Ge atom in a dimer to
form a Sn--Ge heterodimer and releasing that Ge atom
\cite{TomatsuSn,Tomatsu2007Science,HofmannSn}. The system has been studied
closely because Ge--Sn alloys are of interest for optoelectronics, becoming
direct-gap semiconductors at Sn fractions of about 10\%, depending on strain
\cite{Wirths2015}. Using Monte Carlo sampling with DFT energies, Cao et al.\ find that the minority species in Ge--Sn avoids nearest neighbours of its own kind \cite{Cao2020}. At 300~K each Sn in Ge$_{0.75}$Sn$_{0.25}$ has on average $0.33\pm0.02$ Sn nearest neighbours, compared with one in a random alloy. Si$_{0.75}$Ge$_{0.25}$ stays close to random. The authors relate the difference to the 14\% lattice mismatch between Ge and $\alpha$-Sn. At 12.5\% Sn, including short-range order raises the calculated direct gap by about 0.1~eV and improves agreement with experiment. In films with up to 12.4\% Sn, Gencarelli et al.\ resolved second-shell Sn by extended X-ray absorption fine structure (EXAFS). Four samples also admitted a weak first-shell Sn--Sn contribution \cite{Gencarelli2015}. They interpreted the second-shell trend as evidence for a random alloy. Cao et al.\ point out that the scarcity of first-shell Sn pairs is consistent with short-range order \cite{Cao2020}. How dopants are distributed is also an open question for Ga in Ge (Sec.~\ref{sec:introduction}). STM studies aimed at detecting short-range order in Ge--Sn have begun on films grown by MBE at 180~$^\circ$C \cite{Baral2026}.

Tomatsu et al.\ first found dimers of unusual contrast at substrate dimer
positions, amounting to less than 0.01~ML after 0.07~ML of Sn and more
numerous after annealing at 147~$^\circ$C \cite{TomatsuSn}. They then showed
that these are Sn--Ge heterodimers by reversing the tilt of the dimers in a
row with the STM tip at 80~K. The apparent height of the dimer changed when
its tilt was reversed, as it would not for a dimer of two like atoms
\cite{Tomatsu2007Science}. In their DFT calculations the heterodimer with Sn
as the upper atom is 0.06~eV more stable than with Sn as the lower atom, and
Sn in the second layer is less stable by 0.12~eV, so the Sn stays in the
outer layer. The two tilts also differ electronically. With Sn as the lower
atom the heterodimer reflects the electrons of the empty surface band that
runs along the dimer row and sets up standing waves, whereas with Sn as the
upper atom the electrons pass almost unscattered.

Hofmann et al.\ extended this work to four coverages from 0.01 to 1.24~ML,
at which the first Sn layer has closed and some material already occupies a second and, occasionally, a third layer. They compared images at positive and negative sample bias with DFT simulations \cite{HofmannSn}. Of the faint protrusions
at dimer positions, 93\% are heterodimers with Sn as the upper atom, and
images at positive and negative bias are needed to tell them from Ge dimers carrying hydrogen.
The released Ge either reaches a step or joins the adsorbed layer. Relaxed
Sn--Sn and Ge--Ge ad-dimers reproduce the observed shapes equally well, so
the shape of an ad-dimer does not identify its element. Differences in
empty-state brightness between neighbouring ad-dimers are reproduced only
when some of them contain both elements, and Hofmann et al.\ propose such
mixed ad-dimers as one possible configuration that holds the released Ge.
They estimate that the closed first layer contains less than 2\% Ge, and
note that exchange at room temperature limits how abrupt a Ge/Sn interface
can be made \cite{HofmannSn}.

\subsection{Sb and P (group V): exchange on annealing}
\label{sec:groupV}

Sb and P are group-V donors in Ge, the counterparts of the group-III
acceptors Ga and In. Both adsorb as pairs on top of the dimer rows and
replace Ge only on annealing (Table~\ref{tab:adsorbates}).

In the calculations of Mart\'{\i}nez-Guerra et al., whether an Sb dimer replaces a Ge surface dimer depends on where the displaced Ge goes \cite{MartinezGuerra2006SbAdsorption}. With the Ge at its bulk energy, which
corresponds to Ge that has joined a step, exchange is favoured by 0.19~eV
per Sb. With the Ge left on the terrace as an ad-dimer, the adsorbed Sb
dimer is favoured by 0.05~eV per Sb. On Si(001) the same exchange costs
0.13~eV per Sb even with the bulk reference, so the two surfaces differ in
whether exchange is favoured at all \cite{MartinezGuerra2006SbAdsorption}.
Pits and Ge islands form after annealing at 417~$^\circ$C, which
Mart\'{\i}nez-Guerra et al.\ take to support the bulk reference
\cite{MartinezGuerra2006SbAdsorption}. The barrier to exchange was not calculated.

Schofield et al.\ demonstrated the precise placement of individual P donors in Si(001) \cite{Schofield2003}, a step towards donor spin qubits in Si \cite{Zwanenburg}. That work motivated studies of P on Ge(001) \cite{Scappucci2013Review,Mattoni2013}. On Ge(001), P adsorbs intact
as P$_2$ near 100~$^\circ$C. A 1~s anneal at 250~$^\circ$C converts the P
dimers into Ge--P heterodimers, and the displaced Ge forms elongated
islands. At about a monolayer, the displaced Ge reaches single-layer steps during a 5~min anneal at 360~$^\circ$C \cite{Mattoni2013}. For In, terrace islands likewise appear first, and Ge reaches steps after a hotter anneal (Sec.~\ref{sec:in001}). Mattoni et al.\ also find carriers fully activated only
up to about $10^{14}$~cm$^{-2}$ (0.16~ML) and propose that more closely
spaced P forms P--P dimers that donate no electrons \cite{Mattoni2013}. The proposed P--P dimers would be a surface counterpart of the inactive B pairs in diamond (Sec.~\ref{sec:introduction}).

\begin{table*}[tb]
\caption{Ga sites on the (001) and (111) surfaces, with the completed
films in the last row for comparison. \emph{Measured} denotes an
element-specific position or local environment; \emph{supported}
denotes an assignment from
STM, diffraction or calculations without an element-specific
measurement; \emph{modelled} denotes a proposed or calculated
occupation. \emph{Undetermined} means that no site has been assigned. For the surface rows, coordination refers to Ga before overgrowth.}
\label{tab:gasites}
\small
\setlength{\tabcolsep}{5pt}\renewcommand{\arraystretch}{1.17}
\begin{ruledtabular}
\begin{tabular}{L{0.10\textwidth}L{0.31\textwidth}L{0.17\textwidth}L{0.34\textwidth}}
Surface or film & Site and preparation & Principal coordination & Evidence level \\
\hline
Si(001) & Parallel added Ga pair; room-temperature $2\times2$ preparation & Two Si bonds plus the Ga partner & \emph{Measured/supported}: Ga-specific XSW, STM, tunnelling spectroscopy and calculations \cite{QianBedzyk1995,P99,Northrup001} \\
Ge(001) & Parallel pair in ordered half-ML $2\times2$ calculation & Two Ge bonds plus the Ga partner in the relaxed pair model & \emph{Modelled}: DFT for an ordered half-ML layer, with Ga/Si and In/Ge as analogues; dilute site not measured \cite{Takeuchi} \\
Ge(001) & Conversion of double-layer to single-layer steps, about 0.5~ML at 400--500~$^\circ$C & Not determined & \emph{Undetermined}: step change observed by RHEED; Ga site at the steps unknown \cite{Gutakovsky1999} \\
Ge(001) & Annealed $7\times5.5$ mound reconstruction, 0.3--1~ML retained & Threefold sites in the proposed mound model, which holds 0.39~ML Ga & \emph{Modelled}: local Ga occupations from bonding rules; mound morphology observed by STM \cite{P15} \\
Si(111) & $T_4$ adatom in one-third-ML $\sqrt3\times\sqrt3$ lattice & Three Si bonds & \emph{Measured}: XSW; STM \cite{Falta1996,P100} \\
Si(111) & Outer-half substitution in the $6.3\times6.3$ phase, about 0.8~ML & Three Si bonds & \emph{Measured}: XSW \cite{Falta1996,P100} \\
Si(111) & Outer-layer substitution after about half a bilayer of Ge grown at 470~$^\circ$C on the $T_4$ lattice, in small Ga-rich domains & Three bonds & \emph{Measured}: XSW \cite{Falta1996} \\
Ge(111) & Rest-atom substitution in $c(2\times8)$, up to 0.05~ML or from doped crystals & Three Ge bonds & \emph{Supported}: STM at positive and negative sample bias \cite{MolinasMataZegenhagen1993} \\
Ge(111) & Defects pinning Ge-adatom $(2\times2)$ domains & Undetermined & \emph{Modelled}: Ga occupation proposed \cite{P12,P28,MolinasMata1994} \\
Ge(111) & Outer-layer substitution in the $\gamma$ and $\beta$ domains, 0.7--0.8~ML & Three Ge bonds after inward relaxation & \emph{Measured/supported}: Ga height from XSW in the $\gamma$ phase, with DFT and STM of the domains \cite{P09,Bohringer1995Walls,P28,P66} \\
Ga:Ge film on Ge(001) & Substitutional Ga in MBE films grown at 95--150~$^\circ$C & Four bonds & \emph{Measured}: Ga $K$-edge EXAFS; Ga distribution not determined \cite{P32} \\
\end{tabular}
\end{ruledtabular}
\end{table*}

\subsection{Exchange and the fate of displaced Ge}
\label{sec:displacedGe}

The growth temperature of the films, 95--150~$^\circ$C, falls between two
kinds of behaviour among the other adsorbates. Indium, Sb and P remain as
adsorbed pairs at room temperature or near
100~$^\circ$C and replace Ge only on annealing at 200--250~$^\circ$C or
above, whereas Sn replaces some Ge during room-temperature deposition
\cite{FalkenbergIn2002,MartinezGuerra2006SbAdsorption,Mattoni2013,TomatsuSn}.
For Ga the temperature at which exchange begins has been neither measured
nor calculated (Sec.~\ref{sec:gadilute}). The onset of
exchange matters only for Ga that replaces Ge in a completed layer, not for
Ga captured at a step or within an incomplete layer
(Sec.~\ref{sec:growth}). For In, Sb and P, the first evidence of exchange appears together with Ge islands on the terrace. These observations leave open whether the exchange barrier or the mobility of displaced Ge sets the onset temperature.

Where exchange has occurred it leaves both elements unevenly distributed, in
addition to the aggregation described in Sec.~\ref{sec:gadistribution}.
Indium is concentrated in mixed units holding 0.5 In per $1\times1$ cell,
the Ge released by In, Sb and P collects in terrace islands, and the Ge
released by Sn may stay in the adsorbed layer
\cite{FalkenbergIn2002,MartinezGuerra2006SbAdsorption,Mattoni2013,HofmannSn}.

STM alone cannot count the Ga on a Ge(001) surface where exchange has begun.
Displaced Ge appears among the protrusions, as the occupied-state maximum of
the In $3\times4$ unit and, in the configuration proposed by Hofmann et al.,
within the Sn ad-dimers \cite{FalkenbergIn2002,HofmannSn}. Threefold Ga has
STM signatures of its own (Secs.~\ref{sec:gasi001} and \ref{sec:ga111ge}),
but Ga that has gained a fourth neighbour beneath the surface may show none.
An element-specific measurement is therefore needed beside the images.

At room temperature, Sugimoto et al.\ used non-contact atomic force microscopy (AFM) to identify individual atoms on Si(111) \cite{Sugimoto2007}. With the same tip apex, they measured the maximum attractive short-range force over each species and divided it by the force over Si. The ratios varied little from tip to tip and averaged $0.77\pm0.02$ for Sn, $0.59\pm0.03$ for Pb and $0.72\pm0.04$ for In. Calibration on surfaces of known composition allowed them to distinguish Si, Sn and Pb atom by atom in a surface alloy whose topographic contrast alone was ambiguous. Applying the method to Ga on Ge would require its own calibration.
\section{Ga at the growth front}
\label{sec:growth}

\begin{figure}[tbp]
  \centering
  \includegraphics[width=\linewidth]{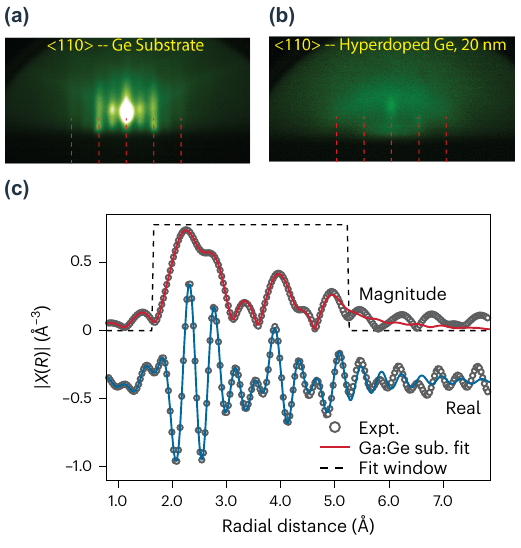}
  \caption{The Ga:Ge growth front and the local environment of incorporated Ga.
(a,b) RHEED along $\langle110\rangle$ from the Ge(001) substrate before deposition and at the end of Ga:Ge growth, before the Si cap. A diffuse ring overlies the weakened crystalline pattern in (b). Steele et al. attribute the ring to a floating Ga layer. (c) Phase-corrected Fourier transform of the Ga $K$-edge EXAFS from the completed film. The upper trace shows the magnitude. The lower trace shows the real part, offset vertically. Circles are measurements, solid lines are the substitutional-Ga fit, and the dashed box marks the fit window. From Supplementary Fig.~S2(a,e) and Fig.~2(c) of Steele et al.~\cite{P32}.}
  \label{fig:fig06}
\end{figure}

MBE is the first method to give superconducting Ga:Ge films without the Ga-rich precipitates found in implanted films \cite{P32}. Implantation amorphises the layer, which must then be regrown, and even the single-crystalline layers obtained by rear-side flash-lamp annealing, with a transition near 0.4~K, cannot exclude Ga clusters below the resolution of electron microscopy \cite{Prucnal}. Chemical vapour
deposition (CVD)
also yields hyperdoped Ge \cite{P64,Ringwala2025}, with up to
$2.7\times10^{20}$~Ga~cm$^{-3}$ in the first report \cite{P64}, more than an
order of magnitude below the Ga concentration of the MBE films. The MBE
films are grown by co-depositing Ga and Ge on Ge(001), and the highest
transition temperatures occur for growth at 95--150~$^\circ$C
\cite[Supplementary Fig.~S3]{P32}. No Ga-rich phase is found in the
completed films. X-ray scattering and Raman spectroscopy detect none,
cross-sectional composition maps show a uniform Ga distribution at their
resolution, and the Ga $K$-edge EXAFS is best described by substitutional Ga [Fig.~\ref{fig:fig06}(c)] \cite{P32}. Ga and Ge differ by one in atomic number and have similar EXAFS backscattering amplitudes and phases. The fit identifies the substitutional environment but cannot reliably separate Ga and Ge in the first neighbour shell. Ga and Ge columns also give nearly the same contrast in high-angle annular dark-field STEM. The larger difference between Sn and Ge (18 in atomic number) makes their contributions distinguishable in EXAFS (Sec.~\ref{sec:groupIV}).

The growers attribute the diffuse ring that appears in the RHEED pattern during growth [Fig.~\ref{fig:fig06}(a,b)] to a layer of Ga floating on the growth front \cite[Supplementary Fig.~S2]{P32}. The diffuse ring suggests disorder in the layer, which acts as a reservoir of Ga during growth. RHEED does not resolve the amount of Ga
in the layer or the way Ga leaves it.

\begin{figure*}[tp]
  \centering
  \includegraphics[width=\linewidth]{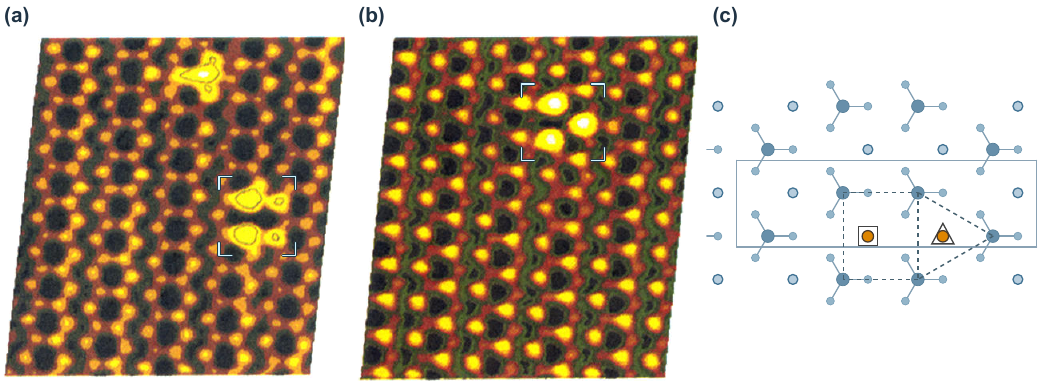}
  \caption{Ga at the two rest-atom sites of Ge(111)-$c(2\times8)$. (a,b) Filled-state STM of Ge crystals doped with approximately $10^{15}$~Ga~cm$^{-3}$ at $-1.0$~V sample bias and 6.7~nA. The corner marks enclose the four bright Ge adatoms around an assigned $R_4$ substitution in (a) and the three around an $R_3$ substitution in (b). (c) Shared site key on the clean reconstruction, in the orientation of Fig.~\ref{fig:fig03}(b). Orange marks the two alternative Ga sites, with a square for $R_4$ and a triangle for $R_3$. Dashed lines connect the adatoms that brighten around each site. Images from Figs.~2(b) and 3(b) of Molinàs-Mata and Zegenhagen~\cite{MolinasMataZegenhagen1993}.}
  \label{fig:fig07}
\end{figure*}

Bulk diffusion of Ga in Ge is negligible at the growth temperatures, a
conclusion that rests on measurements by S\"odervall et al.\ in Ge single
crystals at 554--916~$^\circ$C. Using secondary-ion mass spectrometry to
profile the Ga, they found an Arrhenius law with an activation enthalpy of
$319.5\pm3.4$~kJ~mol$^{-1}$, or 3.31~eV \cite{GaTracer}. These data describe
equilibrium diffusion through crystalline Ge and were taken some
400~$^\circ$C above the growth temperatures. The growth front is not at equilibrium, so these data do not show how far Ga moves at the surface, or in the first layers beneath it, before it is
covered. With an activation enthalpy this large, however, equilibrium diffusion would move buried Ga a negligible distance during growth at 150~$^\circ$C. Since the Ga in
the completed films is substitutional, Ga must be incorporated, and its
distribution set, at or near the growth front.

Whether a dopant stays at an advancing surface or is buried where it sits
depends on the element and the temperature, as early studies of segregation
during MBE established. Jorke studied the segregation of Sb during MBE on
Si(001) \cite{Jorke1988}, and N\"utzel and Abstreiter linked segregation to
surface diffusion in a model that accounts for data on Si, Ge and GaAs
surfaces \cite{Nutzel1996}.
 For Ga in Ge, Kesan et al.\ showed that during MBE on Ge(001) at
450--550~$^\circ$C the Ga segregates to the growth front and is incorporated
only gradually \cite{Kesan}. Strohbeen et al.\ later found that in MBE of
Ga-hyperdoped Ge the segregation depends on the growth conditions
\cite{Strohbeen2023}, whereas Ga-doped Ge grown by CVD from hydrides below
400~$^\circ$C shows no accumulation of Ga at the surface
\cite{Ringwala2025}. P $\delta$-layers formed from PH$_3$ on Ge(001) stay sharp to within the depth resolution of secondary-ion mass spectrometry under Ge overgrown
at up to 460~$^\circ$C \cite{Scappucci2009Encapsulation}.

A layer of Ga carried on a growing surface has precedents in other systems.
On GaN it acts as a surfactant (Sec.~\ref{sec:alphaga}), and adsorbed Ga
reduces defects and roughness during Si homoepitaxy on Si(111) at
300~$^\circ$C \cite{Gallas1996}. Ge deposited at 470~$^\circ$C on the $T_4$
Ga lattice of Si(111) ends up beneath a Ga-terminated surface, where after
about half a bilayer XSW places the Ga in substitutional sites of the outer
layer, within small Ga-rich domains \cite{Falta1996}. This measurement
concerns Ga that stays on top of the growing Ge as a surfactant, on another
surface and at a higher temperature than the superconducting films, so it
does not show how buried Ga is distributed.

Ga could enter the film by capture or by exchange. Ga
captured at a step or within an incomplete layer can take a lattice site as
the layer forms around it. Ga that replaces Ge in a completed layer, and is
then covered, releases a Ge atom to the terrace or a step
(Sec.~\ref{sec:displacedGe}). Ga in the parallel pairs favoured by
calculation would first have to rearrange (Sec.~\ref{sec:gadilute}), whereas capture need not pass through a pair. The
temperature at which Ga begins to exchange with Ge would show whether exchange is open to Ga during growth (Table~\ref{tab:questions}).
\section{Ga on Ge(111)}
\label{sec:ga111}

\subsection{Ga deposited on Ge(111)}
\label{sec:ga111ge}

On Ge(111), deposited Ga can be followed from isolated substitutions to dense surface phases. Both the dilute and dense phases involve threefold Ga in the outer Ge layer (Table~\ref{tab:gasites}). At low coverage Ga replaces rest atoms. Up to 0.05~ML it takes rest-atom
sites of both inequivalent kinds, leaves the Ge adatoms in place and does
not disturb the long-range order of the $c(2\times8)$ reconstruction
\cite{MolinasMataZegenhagen1993}. Figure~\ref{fig:fig07} illustrates the same two substitutions on a Ga-doped crystal. At negative
sample bias, in filled-state images, the adatoms around a substituted site
appear brighter than the others. At positive bias the substituted sites are
not visible, and the surface looks like clean $c(2\times8)$. Molin\`as-Mata
and Zegenhagen attribute the contrast to Ga blocking the transfer of charge
from those adatoms to the rest atom that occurs on clean Ge(111). The assignment rests on STM and is not element-specific. Threefold
Ga is detected here through the contrast of its neighbours, whereas on
Si(001) it is detected through an empty state (Sec.~\ref{sec:gasi001}).

From 0.05 to 0.5~ML the Ga has not been located. Annealing above
300~$^\circ$C gives domains of $(2\times2)$ Ge adatoms, which resemble the
high-temperature phase of clean Ge(111) (Sec.~\ref{sec:clean111}) and in which no Ga has been detected. B{\"o}hringer et al.\ propose that the Ga sits at the defects
that pin these domains \cite{P12,P28,MolinasMata1994}.

The two phases, $\gamma$ near 0.7~ML and $\beta$ near 0.8~ML, form above about 0.5~ML after annealing near or above 500~$^\circ$C. In these dense phases Ga replaces the Ge of the outer layer. XSW places it 0.57~\AA\ below the ideal
outer-layer site in the $\gamma$ phase, close to the 0.49~\AA\ inward
relaxation calculated for substitution. DFT ranks substitution lowest in
energy, followed by substitution with a stacking fault and then the $T_4$
and $H_3$ hollows \cite{P09}. Both are incommensurate, with domains of a local $1\times1$
structure separated by a network of narrow walls, called discommensurations
in these studies, which relieve the compressive stress of the Ga layer by
letting the domains expand laterally
\cite{P09,P28,P66,Bohringer1995Walls}.

The $\beta$ phase has triangular domains of two kinds, with and without a
stacking fault between the Ga layer and the Ge beneath it. XSW places about
40\% of the Ga in the faulted domains, and STM shows them as half of the
triangular domains. XSW counts atoms in each stacking environment, whereas STM counts domains, which can have different areas. B{\"o}hringer et al.\ propose that the fault allows
walls free of dangling bonds, which would drive the change from $\gamma$ to
$\beta$ \cite{P09,Bohringer1995Walls}. At higher coverage a second Ga layer
occupies $H_3$ sites \cite{P09}.

Ga on Si(111) also substitutes into the outer half of the bilayer near
0.8~ML, in the incommensurate $6.3\times6.3$ phase \cite{Falta1996,P100}.
The two surfaces differ at one-third of a monolayer, where only Si(111) forms a Ga adatom lattice (Sec.~\ref{sec:sites111}), but both take dense Ga into the outer layer. P, by
contrast, forms trimers in place of the Ge adatoms on Ge(111) and does not
enter the outer layer \cite{Vitali2001}. Where the Ge displaced by Ga goes
on Ge(111) is an open question.

\subsection{Ga at the surfaces of doped crystals}
\label{sec:ga111doped}

Ga-doped crystals show where Ga appears at a Ge(111) surface without deposition and offer a first substrate for the Sn experiment in Sec.~\ref{sec:snpb111}. Ga atoms from the bulk appear at the same rest-atom sites as on dosed surfaces \cite{MolinasMataZegenhagen1993}. They are
seen after sputtering and annealing at 650~$^\circ$C on crystals with about
$10^{15}$ and about $10^{19}$~Ga~cm$^{-3}$, more often at the higher
concentration. Figure~\ref{fig:fig07}(a,b) shows substitutions on the lightly doped crystals ($\sim10^{15}$~Ga~cm$^{-3}$). At this concentration
random occupation at the bulk Ga fraction would give only one Ga atom
per 25~$\mu$m$^2$ among the rest sites, which occupy 0.25~ML
(Table~\ref{tab:scales}). Finding substitutions in these nanometre-scale images is evidence that preparation enriches the outer layer in Ga. The density of Ga in the outer layer has not
been measured against the bulk concentration (Sec.~\ref{sec:outlook}).

Molin\`as-Mata et al.\ compared vicinal surfaces of crystals with about
$10^{19}$~Ga~cm$^{-3}$, which keep the $c(2\times8)$ reconstruction, with
those of Czochralski crystals of the highest Ga content,
$2.9\times10^{20}$~Ga~cm$^{-3}$ with $2\times10^{20}$~holes~cm$^{-3}$
\cite{MolinasMata1993Doped}. The latter also contain more than
$10^7$~dislocations~cm$^{-2}$ and hold part of their Ga in inclusions. On
their surfaces misoriented by 1$^\circ$ or 2$^\circ$, each terrace forms an
incommensurate honeycomb of $(2\times2)$ domains separated by
$c(4\times2)$ walls at room temperature, the structure proposed for the
high-temperature phase of clean Ge(111). Their well-oriented
surfaces keep the $c(2\times8)$. The authors conclude that the
misorientation and the dislocations are both needed, and suggest that
strain from the inclusions produces the dislocations and that Ga alters the
energy of the wall crossings.

\subsection{Sn/Ge(111): correlated adatom lattices and a possible route to hole doping}
\label{sec:snpb111}

\begin{figure*}[tp]
  \centering
  \includegraphics[width=\linewidth]{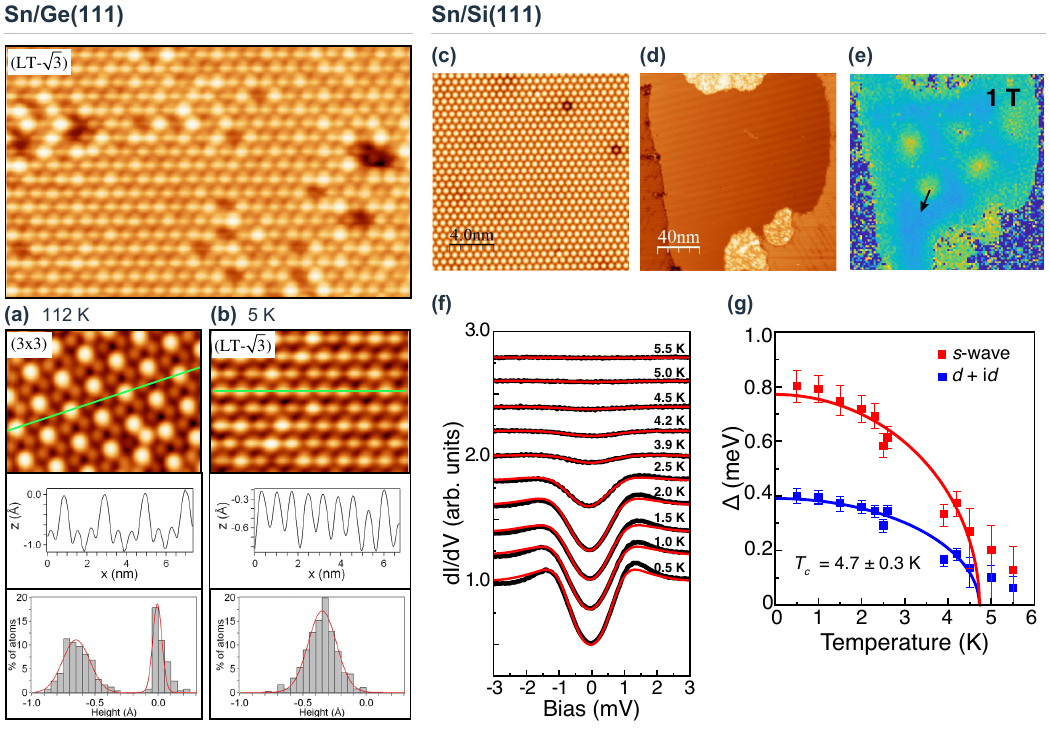}
  \caption{Sn adatom lattices on Ge(111) and hole-doped Si(111). On Ge, the $18\times11$~nm$^2$ overview shows the low-temperature $\sqrt3$ phase reported by Cortés et al. at 5~K. Beneath it, (a) and (b) compare $7\times5$~nm$^2$ fields of the $(3\times3)$ phase at 112~K ($-1.0$~V) and the $\sqrt3$ phase at 5~K ($-1.4$~V). The apparent-height profile along each green line and the distribution of adatom heights are shown below the corresponding image. All three images were taken at 1.0~nA. From Fig.~2 of Cortés et al.~\cite{Cortes2006}. (c) Atomic-resolution STM of the triangular Sn lattice on Si(111) ($+1.0$~V, 1.0~nA). (d,e) Topograph and zero-bias conductance map of the same Sn domain. The bright patches in (e) are vortex cores at 0.5~K and 1~T. (f) Zero-field tunnelling spectra at different temperatures, normalised to the normal-state spectra and offset vertically. Red curves are $s$-wave Dynes fits. (g) Gap magnitudes extracted using $s$-wave and chiral $d$-wave fits. The curves are mean-field fits to $\Delta(T)$, giving $T_c=4.7\pm0.3$~K. Panels (c--g) are from Figs.~1(c), 4(a,b) and 3(a,b) of Wu et al.~\cite{Wu2020}.}
  \label{fig:fig08}
\end{figure*}

The Sn lattice on Ge(111) is metallic with a $(3\times3)$ distortion down to about 60~K, and what it becomes at lower temperature is disputed. It has been studied
since the 1990s by STM, LEED, photoemission, surface X-ray diffraction and
calculation. To the author's knowledge, no transport measurement of this Sn lattice has been reported.

Sn occupies $T_4$ sites at
one-third of a monolayer (Sec.~\ref{sec:sites111}), and at room temperature
the lattice appears $\sqrt3$ because the adatoms fluctuate vertically
\cite{Ortega2002,Halbig2019}. Below about 210~K an order--disorder
transition gives the $(3\times3)$ phase, with one adatom in three raised
above the other two by 0.26--0.38~\AA\
\cite{Zhang1999,Bunk1999SnGe,Shirasawa2010,Halbig2019}. Figure~\ref{fig:fig08}(a) shows the $(3\times3)$ contrast at 112~K and the apparent height along the marked line. Defects occupy
$3\pm1$\% of the adatom sites, mostly Ge in place of Sn, and pin the
$(3\times3)$ order. By 105~K STM resolves ordered $(3\times3)$ domains in which the defects align on a honeycomb sublattice. The mean defect spacing is about 45~\AA\ \cite{Melechko2000}.

Cort\'es et al.\ find that near 60~K the lattice becomes a charge-ordered $(3\times3)$ phase, with three inequivalent Sn heights and a gap of about 40~meV in tunnelling spectra \cite{Cortes2013}. Below about 30~K they report that a flat $\sqrt3$ phase forms gradually [Fig.~\ref{fig:fig08}(b)], with a gap of about 100~meV, which they interpret as a Mott insulator \cite{Cortes2006,Cortes2013}. Domains of neighbouring phases coexist near 60~K and near 20~K. Colonna et al.\ and Morikawa et al.\ instead image a $(3\times3)$ phase at 4--10~K that remains metallic \cite{Colonna2008,Morikawa2008}, which Cort\'es et al.\ identify with their charge-ordered phase. A $(3\times3)$ periodicity alone therefore does not show that the lattice is metallic. Shirasawa et al.\ find by LEED the same $(3\times3)$ structure at 14~K as at 80~K \cite{Shirasawa2010}, whereas Cort\'es et al.\ see only $\sqrt3$ spots by 49~K in LEED but see the flat phase emerge below about 30~K in STM \cite{Cortes2013}. The measurements themselves may contribute to the discrepancy, since the tunnelling current and the LEED electron beam can both alter the $(3\times3)$ order \cite{Colonna2008,Shirasawa2010,Cortes2013}. A Slater
insulator \cite{Lee2013}, a Peierls-like distortion mediated by substitutional Ge defects
\cite{Chiang2002} and strong electron--phonon coupling have also been proposed. For the $(3\times3)$ phase, Nair et al.\ obtain an electron--phonon coupling constant $\lambda=1.45\pm0.1$ from photoemission and 1.3 by calculation \cite{Nair2023}, and attribute part of the change in the spectra down to 5~K to coupling that grows as the temperature falls \cite{Nair2024}. Marchetti et al.\ calculate electron--phonon coupling for Sn and Pb $\sqrt3$ lattices on Si(111) and SiC(0001). For all four they find the coupling too weak to drive conventional superconductivity. Their coupling calculations do not cover Ge(111) \cite{Marchetti2025}. The half-filled band is narrow, with an on-site repulsion comparable to its width. Calculations place the flat $\sqrt3$ phase,
the $(3\times3)$ phase and a charge-ordered state within about 10~meV per adatom of one another \cite{ProfetaTosatti2007,Cortes2013}.

On Si(111) the same triangular adatom lattice [Fig.~\ref{fig:fig08}(c)] has been doped through the substrate. Profeta
and Tosatti proposed in 2007 that it might superconduct away from half
filling \cite{ProfetaTosatti2007}. The high-temperature anneal that cleans
heavily B-doped Si brings B towards the surface before Sn is deposited
\cite{Ming2017,Ming2023,Wu2020}. The Sn lattice grown on that surface has a
hole doping of order 10\% \cite{Ming2017,Ming2017Nano} and is
superconducting \cite{Wu2020}, with evidence for chiral pairing
\cite{Ming2023,Wu2026Chiral}. Figure~\ref{fig:fig08}(d,e) shows the Sn domain and its vortex cores in a 1~T field. From the temperature dependence of the tunnelling gap [Fig.~\ref{fig:fig08}(f,g)], Wu et al.\ obtain $T_c=4.7\pm0.3$~K \cite{Wu2020}.

The equivalent experiment has not been tried on Ge(111). The studies that
report their substrates used lightly doped Ge, with resistivities between
about 0.1 and 50~$\Omega$~cm, and the $(3\times3)$ phase appears on n-type
and p-type crystals alike \cite{Cortes2006,Morikawa2008,Shirasawa2010}. Ga
appears in the outer layer of Ga-doped Ge(111) crystals after annealing
(Sec.~\ref{sec:ga111doped}), so Ga could plausibly dope a Sn
lattice grown on such a surface, or on a hyperdoped substrate, as B does on
Si(111).

The outcome would depend on where the Ga sits, as a simple electron count
shows (Sec.~\ref{sec:counting}). Ga in the adatom lattice in place of
Sn would be threefold and contribute no dangling-bond electron, so in this count it acts as a vacancy in the Sn band, not as a dopant. Ga in the outer Ge layer, the
site it takes on bare Ge(111), would lie among the three atoms to which each
Sn adatom bonds \cite{Bunk1999SnGe}, where it would have a fourth neighbour and
could take an electron from the Sn band directly. Ga deeper in the crystal
is an ordinary acceptor and changes the filling only through the space-charge layer beneath the surface. For Sn/Si(111), Iannetti et al.\ calculate that B at the usual subsurface site depletes the dangling bond of the Sn above it \cite{Iannetti2025}. In a local electron count this removes one electron and its site from the low-energy band, leaving the remaining sites at half filling. The authors propose that deeper B helps make the lattice metallic. Ga near the surface at a density
comparable to the doping reached on
Si(111) would also be at least as numerous as the Ge defects that pin the
$(3\times3)$ order, so doping and pinning may not be separable.

\begin{figure*}[tp]
  \centering
  \includegraphics[width=\linewidth]{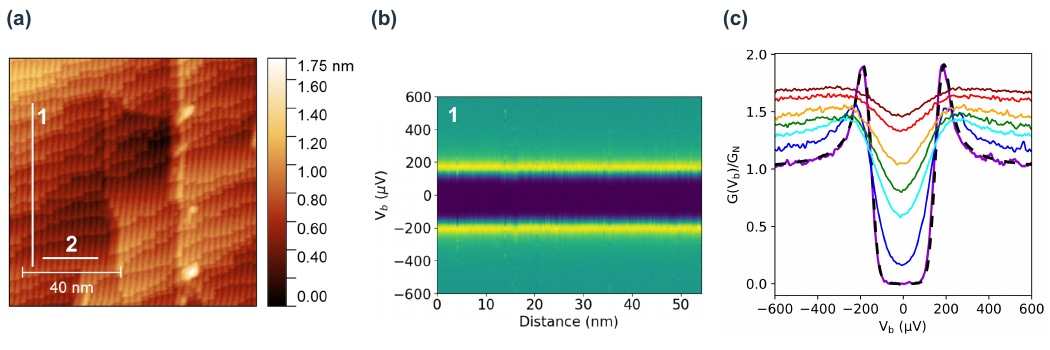}
  \caption{Tunnelling spectroscopy of $\alpha$-Ga. (a) Constant-current STM of the (112) facet ($+100$~mV, 50~pA). Line 1 crosses steps and line 2 crosses the reconstruction stripes. (b) Differential conductance along line 1 in (a) at base temperature.  The coherence peaks stay at the same bias across the steps. Dark blue to yellow spans zero to approximately twice the normal-state conductance. (c) Spectra taken with a normal-metal tip at cryostat temperatures from 50 to 970~mK, offset vertically. The dashed line is a Bardeen--Cooper--Schrieffer fit to the lowest-temperature spectrum. From Figs.~3(a), 3(c) and 2(a) of Fohn et al.~\cite{Fohn2024}. The corresponding measurement along line 2 is their Fig.~3(d).}
  \label{fig:fig09}
\end{figure*}
\section{Elemental and two-dimensional Ga}
\label{sec:alphaga}

\begin{table*}[tb]
\caption{Open questions for Ga at Ge surfaces and in the films, with the
measurements and calculations that would answer them. The last row is an
emerging question that links the films to the Sn adatom lattice on
Ge(111). APT, atom-probe tomography.}
\label{tab:questions}
\small
\setlength{\tabcolsep}{5pt}\renewcommand{\arraystretch}{1.2}
\begin{ruledtabular}
\begin{tabular}{L{0.30\textwidth}L{0.63\textwidth}}
Open question & What would answer it \\
\hline
Where does dilute Ga sit on Ge(001) at 95--150~$^\circ$C, and when does it
exchange with Ge? & \emph{Surface.} STM registered to the substrate lattice below 0.1~ML at the
growth temperatures, with element-specific measurements such as XSW, core-level photoemission or non-contact AFM force spectroscopy,
including the onset of exchange and where the displaced Ge goes.
\emph{Calculation.} Isolated pairs, row-top and substituted sites, and
exchange with the displaced Ge on the terrace or at a step
\cite{Takeuchi,QianBedzyk1995,MartinezGuerra2006SbAdsorption}. \\
How does the floating layer pass Ga into the film? & \emph{Surface.} Ga
content and lateral distribution of growth surfaces interrupted before and
after one or a few Ge layers \cite{P32,Falta1996}. \\
How is Ga distributed in the films? & \emph{Film.} Composition variation
across a film by APT, and the frequency of Ga--Ga nearest neighbours at a
given composition, for example by resonant diffuse scattering near the Ga $K$ edge in thicker films. \emph{Calculation.} Energies and bands of regular,
random and locally Ga-rich distributions in cells large enough to compare
them \cite{P32,Heera2014,Gault2022,Lebegue2009}. \\
What structures does Ga form on Ge at higher temperatures and coverages? &
\emph{Surface.} Element-specific measurements of the Ge(001) phases formed
above about 400~$^\circ$C and of Ge(111) between 0.05 and 0.5~ML, so far
examined mainly by STM, LEED and RHEED. \emph{Calculation.} Relaxed
energies of the proposed structures
\cite{Ji1997,Gutakovsky1999,P15,P12}. \\
Emerging: can Ga hole-dope the Sn lattice on Ge(111)? & \emph{Surface.} Ga
site, band filling and order of Sn lattices on Ga-doped Ge(111) crystals or
hyperdoped films, and the outer-layer Ga density of doped crystals.
\emph{Calculation.} Electron counts and bands of Ga in the adatom lattice,
in the outer Ge layer and deeper, as for B beneath Sn/Si(111)
\cite{MolinasMataZegenhagen1993,Ming2017,Wu2020,Iannetti2025}. \\
\end{tabular}
\end{ruledtabular}
\end{table*}

Elemental Ga is a superconductor whose transition temperature depends
strongly on its phase. The stable phase, $\alpha$-Ga, has its transition at
1.08~K \cite{Fohn2024,Campanini2018}. Metastable $\beta$-Ga has its
transition at 6.04~K \cite{Campanini2018}, and amorphous Ga films condensed at 4~K become superconducting at 8.4~K \cite{Buckel1954,Gorlach1982}. Ga-rich clusters or inclusions
occur in implanted films and in heavily doped crystals
(Secs.~\ref{sec:introduction} and \ref{sec:ga111doped}), and the higher
transition temperatures of $\beta$ and amorphous Ga could account for some
of the superconductivity observed in implanted material. In Ga-implanted Si
capped with SiO$_2$, a transition at 7~K follows the precipitation of Ga
beneath the oxide \cite{SkrotzkiSiGa2010}. In Ge prepared in the same way,
a transition at 6~K fell below 1~K once the amorphous Ga-rich layer at the
oxide interface was etched away \cite{Fiedler2012}, and in other implanted Ga:Ge films the superconductivity is controlled by the coupling between Ga
precipitates \cite{Sardashti}. Segregated $\alpha$-Ga is accordingly one of
the models against which the EXAFS of the MBE films was tested \cite{P32}.

The structure and surfaces of $\alpha$-Ga have been studied by STM, X-ray
diffraction, LEED and calculation
\cite{Bernasconi1993,ZugerDurig1992,Walko1998,More2003}, and its
superconducting state has now been measured on the atomic scale. Fohn et
al.\ imaged atomic terraces and the (112) facet
[Fig.~\ref{fig:fig09}(a)] and measured the gap by tunnelling spectroscopy
at a cryostat base temperature of 50~mK. The STM thermometer read 115~mK, and a fit to the base-temperature spectrum gave an effective tip temperature of about 140~mK \cite{Fohn2024}. The spectra follow Bardeen--Cooper--Schrieffer theory
[Fig.~\ref{fig:fig09}(c)] and are highly homogeneous across the surface,
including at atomic defects and step edges [Fig.~\ref{fig:fig09}(b)], an observation that Fohn et al.\ cite as evidence against an earlier suggestion of topological superconductivity based on magnetotransport \cite{ChenGa2018,Fohn2024}. That
uniformity is a
benchmark against which tunnelling spectra of a Ga:Ge surface, where the Ga
distribution is still to be determined, could be compared.

Between the dilute and the elemental limits lie Ga layers one or two atoms
thick at semiconductor surfaces and interfaces, and several of these layers are superconducting. On GaN(0001) a Ga bilayer forms on the growing surface during MBE and
acts as a surfactant, with about 2.5~ML retained over a wide range of Ga
fluxes \cite{Northrup2000GaN,Mula2001,Adelmann2003}. The layer is metallic
\cite{Sun2010GaN}, and two-monolayer Ga films grown on GaN(0001) become
superconducting near 5.4~K \cite{Zhang2015Ga}. Between epitaxial graphene
and SiC, intercalated Ga forms atomically thin layers bonded covalently to
the SiC below and not bonded to the graphene above, with a transition near
4~K \cite{Briggs2020}. Prepared under ultrahigh vacuum, the intercalated
layers show striped and hexagonal moir\'e superstructures and coexisting
phases, among them one that Pompei et al.\ attribute to Ga\,III, a dense
phase of bulk Ga stable only at high pressure and temperature
\cite{Pompei2025,Bernasconi1993}. Among the Ga phases considered here at semiconductor interfaces, these layers are the most Ga-rich. The floating Ga at the Ge growth front
(Sec.~\ref{sec:growth}) and the dense Ga phases of Ge(111)
(Sec.~\ref{sec:ga111ge}) lie between these Ga-rich layers and the dilute
substitutional Ga of the films.
\section{Summary and outlook}
\label{sec:outlook}

The adsorption site of dilute Ga on Ge(001) has yet to be determined at
95--150~$^\circ$C, the temperatures that yield superconducting films.
At these temperatures an adsorbed Ga atom can gain its fourth neighbour at the growth front, where nearby Ga atoms may be brought together or kept apart. It is still unclear how Ga passes from the floating layer into the film, by capture during
overgrowth or by exchange with Ge (Sec.~\ref{sec:growth}).

In Josephson junctions grown from the Ga:Ge/\allowbreak Si/\allowbreak Ga:Ge trilayer of Fig.~\ref{fig:fig01}
\cite{Knudsen2026}, the upper Ga:Ge layer begins on the thin Si layer, so Ga
at that growth front first meets a Si(001) surface under tensile strain from the Ga:Ge beneath. The Ga site measured on Si(001) (Sec.~\ref{sec:gasi001}) is therefore a
starting point for that interface as well.

One way to follow incorporation would be to interrupt growth before and
after adding a Ge layer (Table~\ref{tab:questions}). STM could map pairs and
rows on the exposed surface. Once exchange begins, both Ga and displaced
Ge can appear in the images, so an element-specific measurement would be needed to locate Ga (Sec.~\ref{sec:displacedGe}). XSW could locate Ga relative to the lattice planes, and high-resolution core-level photoemission could distinguish chemical environments through their core-level shifts. Both are usually performed with synchrotron radiation. Non-contact AFM force spectroscopy offers a local measurement, as demonstrated for individual atoms on Si(111) \cite{Sugimoto2007}.

Atom-probe tomography has resolved Ga-rich precipitates in implanted films
\cite{Heera2014}. In the epitaxial films it could test for nanometre-scale
composition variations, though individual Ga--Ga nearest neighbours are
generally beyond its resolution \cite{Gault2022}. If thicker films can be grown, resonant diffuse X-ray scattering near the Ga $K$ edge could use the enhanced Ga--Ge contrast to measure Ga--Ga pair correlations \cite{IceSparks1999}.

On Ge(111), the density of Ga in the outer layer of a doped crystal has
yet to be measured against the bulk concentration
(Sec.~\ref{sec:ga111doped}). Ga:Ge(111) films have not been grown, but a superconducting Ga:Ge
film beneath a Sn lattice could supply holes to the Sn band and, below the film transition, couple the lattice to a superconductor by proximity. The Sn lattice is ordered by annealing near 290~$^\circ$C (Sec.~\ref{sec:sites111}), whereas growth above 150~$^\circ$C already weakens the superconductivity of the films, with a loss of holes that the growers attribute to Ga segregation \cite{P32}. A first experiment on doped bulk crystals could establish whether Ga
supplies holes to the Sn band, pins the $(3\times3)$ order, or does both.

Hyperdoping already has applications in sub-bandgap infrared detection in
Si \cite{Sher2023}. In Ge, the first crystalline Ga:Ge/Si/Ga:Ge Josephson
junctions \cite{Knudsen2026} bring the question of dopant distribution into
superconducting device physics. The growth fluxes give the expected Ga content of the films. How often two Ga atoms are nearest neighbours remains an open question.
Calculations so far place one Ga atom in a periodic cell of eight or sixteen atoms \cite{P32,Lebegue2009,Prucnal}, which keeps the calculation tractable but excludes Ga--Ga nearest neighbours. With measurements of Ga pair frequencies, larger-cell calculations could
compare the bands and electron--phonon coupling of ordered and disordered
films at the same Ga concentration. If growth can vary that order at fixed Ga content, measurements of the film transition temperature and the junction critical current could test its influence on superconductivity.

\bibliographystyle{apsrev4-2}
\bibliography{references}
\end{document}